\documentstyle[12pt]{article}

\newcommand{\sect}[1]{\setcounter{equation}{0}\section{#1}}

\csname @addtoreset\endcsname{equation}{section}

\def\bseq{\begin{subequation}}  % = 1a 1b
\def\eseq{\end{subequation}}
\def\bsea{\begin{subeqnarray}}  % = 1.1a 1.1b
\def\esea{\end{subeqnarray}}
\newcommand{\bbox}{\lower.2ex\hbox{$\Box$}}

\newcommand{\beq}{\begin{equation}}
\newcommand{\eeq}{\end{equation}}
\newcommand{\bea}{\begin{eqnarray}}
\newcommand{\eea}{\end{eqnarray}}
\newcommand{\ena}{\end{eqnarray}}

\renewcommand{\a}{\alpha}
\renewcommand{\b}{\beta}

\renewcommand{\d}{\delta}

\newcommand{\pa}{\partial}
\newcommand{\g}{\gamma}
\newcommand{\G}{\Gamma}

\newcommand{\D}{\Delta}
\newcommand{\e}{\epsilon}

\renewcommand{\l}{\lambda}
\renewcommand{\L}{\Lambda}
\newcommand{\m}{\mu}

\newcommand{\n}{\nu}

\renewcommand{\P}{\Pi}

\newcommand{\s}{\sigma}
\renewcommand{\S}{\Sigma}

\renewcommand{\O}{\Omega}

\newcommand{\Db}{\bar{D}}

\newcommand{\Sigmab}{\bar{\Sigma}}
\newcommand{\Phib}{\bar{\Phi}}

\newcommand{\mb}{\bar{\mu}}

\begin{document}

\begin{titlepage}
\begin{flushright} IFUM--612--FT\\VUB/TENA/98/2 
%\\ hep-th/98yyxxx
\end{flushright}
\vfill
\begin{center}
{\LARGE\bf Geometry and beta-functions for $N=2$ matter models in two dimensions}\\
\vskip 27.mm  \large
{\bf   Silvia Penati $^1$, Andrea Refolli $^1$,\\
Alexander Sevrin $^{2}$ and  Daniela Zanon $^1$ } \\
\vskip 0.5cm
{\small
 $^1$ Dipartimento di Fisica dell'Universit\`a di Milano and\\
INFN, Sezione di Milano, via Celoria 16,
I-20133 Milano, Italy}\\
\vskip 0.3cm
{\small
$^2$ Theoretische Natuurkunde, Vrije Universiteit Brussel\\ 
Pleinlaan 2, B-1050 Brussel, Belgium }
\end{center}
\vfill

\begin{center}
{\bf ABSTRACT}
\end{center}
\begin{quote}
We study renormalizable nonlinear $\s$-models in two dimensions with
$N=2$ supersymmetry described in superspace in terms of chiral and
complex linear superfields. The geometrical structure of the underlying
manifold is investigated and the one-loop divergent contribution to
the effective action  is computed.
The condition of vanishing $\b$-function allows to identify a class
of models which satisfy this requirement and possess $N=4$ 
supersymmetry. 

\vfill     
\vskip 5.mm
 \hrule width 5.cm
\vskip 2.mm
{\small
\noindent e-mail: penati@mi.infn.it\\
\noindent e-mail: refolli@mi.infn.it\\
\noindent e-mail: asevrin@tena4.vub.ac.be\\
\noindent e-mail: zanon@mi.infn.it}
%\vfill      %\hrule width 5.cm
%\vskip 2.mm
%{\small
%\noindent $^\dagger$ Onderzoeksdirecteur FWO, Belgium }
\end{quote}
\begin{flushleft}
March 1998
\end{flushleft}
\end{titlepage}

%\tableofcontents{}
\section{Introduction}

Nonlinear $\s$-models with $N=2$ supersymmetry
have been extensively studied in the past.
A complete off-shell description
of these systems has been presented in terms of chiral, twisted chiral 
and semi-chiral superfields \cite{SevrinTroost}. Besides these types of
models there exists another class of theories that seems to be 
suitable for a 
description of supersymmetric extensions of the low-energy QCD
effective action \cite{somejim,Jim,Jimetal}.
These models
are generically constructed by means of two
kinds of superfields, i.e. the chiral superfield $\Phi$
and the complex linear superfield $\Sigma$. 
Both formulations provide a description of the
scalar multiplet, in its minimal and nonminimal version respectively.
A duality transformation connects the two multiplets, so that a model
defined in terms of e.g. chiral superfields is essentially equivalent
to the dual one written in terms of linear superfields. This duality, 
implemented
at the classical level by a Legendre transform on the action, has
been studied at the one-loop level and proven to be maintained
\cite{PRVPZ}.

The lagrangian of a renormalizable nonlinear $\s$-model 
in $N=2$, $2$-dimensional superspace is given by a real, nonderivative 
function of the superfields, the potential. 
In the chiral description one can interpret
the superfields as holomorphic coordinates on a K\"ahler manifold
with metric simply given by the second, holomorphic-antiholomorphic 
derivatives of the potential. The geometrical quantities
are then constructed in standard manner and the covariant structure of 
physical objects is easily implemented.
The problem of finding the corresponding geometrical interpretation
for the nonminimal formulation of the theory has been addressed in
ref. \cite{Jim}, where the analysis has been extended
to the generalized $\s$-model whose potential is a function of
chiral and linear superfields.
Here we try to learn more: for such a theory first we discuss the
geometrical properties and find that, 
once the auxiliary $N=1$ components of the complex linear superfields are 
set on-shell, the underlying manifold is K\"ahler. On one hand this 
is not surprising since, using the on-shell condition, we eliminate 
the extra auxiliary degrees of freedom that distinguish the nonminimal
multiplet from the minimal one. On the other hand the result is not a 
priori expected since, as argued in \cite{deogates}, eliminating the 
auxiliary fields of the linear multiplet
modifies the quadratic action of the physical 
components in a nontrivial way.
Indeed we show that this nontrivial behavior of the auxiliary fields 
leads to a $N=1$ or
component action which is the Legendre transform of the chiral one. Thus
the two models are connected by a duality transformation, at least classically
and with the auxiliary fields on-shell.
Actually the equivalence between the two 
formulations is more stringent than this: we
have computed in $N=2$ superspace the one-loop $\b$-function for the mixed, 
chiral plus linear
theory, and we have
found that the quantum, off-shell result is consistent with the above mentioned
duality between the chiral and the complex linear superfields.

Finally we have studied the condition of vanishing $\b$-function and 
identified a class of models which satisfy this requirement 
and possess $N=4$ supersymmetry.
These issues are presented in detail in the following sections.

\section{The geometry}

The chiral $N=2$ nonlinear $\s$-model is described by the superspace
action
\beq
S=\int d^2x~d^4\theta~K(\Phi,\Phib)
\label{chiralaction}
\eeq
with $\Phi^\m$, $\Phib^{\bar{\m}}$, $\m,\bar{\m}=1,\dots,n$, 
satisfying the chirality constraints $\Db_\a \Phi^\m=0$,
$D_\a \Phib^{\bar{\m}}=0$. It is well known \cite{zumino} that 
a simple geometrical interpretation emerges: $K$ represents the K\"ahler 
potential of an underlying manifold  with complex coordinates
$\Phi^\m$, $\Phib^{\bar{\m}}$, whose metric is given by
\beq
g_{\m\bar{\m}}=\frac{\pa^2 K}{\pa\Phi^\m \pa \Phib^{\bar{\m}}}
\label{Kahlermetric}
\eeq
Here and in the following we introduce the notation
\beq
K_{\m\dots\bar{\m}\dots}=
\frac{\pa}{\pa \Phi^\m}\dots \frac{\pa}{\pa \Phib^{\bar{\m}}}\dots 
K(\Phi,\bar{\Phi})
\eeq
so that
\beq
g_{\m\bar{\m}}=K_{\m\mb}
\eeq
In this complex basis the standard quantities of riemannian
geometry become very simple. The only non-vanishing components of
the connection are
\beq
\G^\m_{~\n\rho}=g^{\m\mb}\frac{\pa}{\pa\Phi^{\n}}g_{\rho\mb}
\qquad\qquad \G^{\mb}_{~\bar{\n}\bar{\rho}}=(\G^\m_{~\n\rho})^*
\label{connection}
\eeq
The Riemann tensor can be written
as
\beq
R_{\m\mb\n\bar{\n}}=K_{\m\mb\n\bar{\n}}-K^{\rho\bar{\rho}}
K_{\m\n\bar{\rho}}K_{\rho\mb\bar{\n}}
\label{Riemann}
\eeq
where $K^{\rho\bar{\rho}}$ is the inverse of the metric in
(\ref{Kahlermetric}). Finally the Ricci tensor takes the form
\beq
R_{\m\mb} =\frac{\pa}{\pa \Phi^\m}\frac{\pa}{\pa \Phib^{\mb}}~ 
{\rm ln~ det~ g_{\n\bar{\n}}}
\label{Ricci}
\eeq

The Ka\"hlerian nature of the geometry is most easily recognized 
once the theory is rewritten using the $N=1$ formalism, so that
the relevant complex structure is immediately identified. 
We briefly review the chiral example in order to compare 
later on with the corresponding results in the less familiar complex
linear setting.

\vspace{0.8cm}
We expand the $N=2$ chiral superfield $\Phi$ in $N=1$ superfield
components (we use the notations given in Appendix A)
\beq
\Phi(\theta_1,\theta_2)=\phi(\theta_1) +\frac{1}{2}\theta^\a_2 
\psi_\a(\theta_1)+\frac{1}{4}\theta^2_2F(\theta_1)
\label{Phiexpansion}
\eeq
where
\beq
\phi=\Phi|_{\theta_2=0}\qquad\qquad\psi_\a=D_{2\a}\Phi|_{\theta_2=0}
\qquad\qquad F=-D^2_2\Phi|_{\theta_2=0}
\label{chiralcomponents}
\eeq
The constraint equation $\Db_\a \Phi=0$, which obviously implies 
$\Db^2 \Phi=0$, allows to solve for $\psi_\a$ and $F$ in terms
of $\phi$. We have
\beq
D_{2\a}\Phi=D_{1\a}\Phi \qquad \qquad D^2_2\Phi=-D^2_1\Phi+
D^\a_1 D_{2\a} \Phi=D^2_1\Phi
\eeq
so that in $N=1$ language we obtain
\beq
\psi_\a=D_{1\a}\phi\qquad\qquad F=-D^2_1\phi
\label{chiralsolution}
\eeq
In the same way from the antichirality constraints, i.e.
$D_\a \Phib=0$ and $D^2 \Phib=0$ one has
\beq
\bar{\psi}_\a=-D_{1\a}\bar{\phi}\qquad\qquad \bar{F}=-D^2_1\bar{\phi}
\label{antichiralsolution}
\eeq
Correspondingly the action in (\ref{chiralaction}) becomes
\beq
S=\frac{1}{4}\int d^2x~d^2\theta_1~d^2\theta_2~K(\Phi,\Phib)
=\frac{1}{2}\int d^2x~d^2\theta_1~K_{\m\bar{\n}}(\phi,\bar{\phi})
D_{1\a}\phi^\m
D_1^\a\bar{\phi}^{\bar{\n}}
\label{N1chiralaction}
\eeq
It is manifestly invariant under the first supersymmetry 
transformation $\d \phi^\m=i\e^\a Q_{1\a} \phi^\m$, 
$\d \bar{\phi}^{\mb}=i\e^\a Q_{1\a} \bar{\phi}^{\mb}$, while
the invariance under the second supersymmetry 
\bea
&&\d \phi^\m=i\eta^\a Q_{2\a} \Phi^\m |_{\theta_2=0}=-
\eta^\a \psi_\a^\m \nonumber\\
&&\d \bar{\phi}^{\mb}=i\eta^\a Q_{2\a} \Phib^{\mb}|_{\theta_2=0}=-
\eta^\a \bar{\psi}_\a^{\mb}
\label{SStwo}
\eea
leads to the determination of the complex structure $J^\m_{~\n}$,
$\bar{J}^{\mb}_{~\bar{\n}}$. Indeed
from (\ref{SStwo}) and (\ref{chiralsolution}) it follows
\bea
&&\d \phi^\m=-\eta^\a D_{1\a} \phi^\m\equiv
i\eta^\a J^\m_{~\n} D_{1\a} \phi^\n\nonumber\\
&&\d \bar{\phi}^{\bar{\m}}=\eta^\a D_{1\a} \bar{\phi}^{\bar{\m}}\equiv
i\eta^\a \bar{J}^{\bar{\m}}_{~\bar{\n}} D_{1\a} \bar{\phi}^{\bar{\n}}
\label{complexstructure}
\eea
Thus we identify $J^\m_{~\n}=i\d^\m_{~\n}$ and 
$J^{\bar{\m}}_{~\bar{\n}}=-i\d^{\bar{\m}}_{~\bar{\n}}$. Using the complex 
holomorphic basis
$(\phi^\m, \bar{\phi}^{\bar{\m}})$ the complex structure can be written
in standard form
\beq
J=i\left( \matrix{\bf{1} & \bf{0}\cr
\bf{0} & -\bf{1}}\right)
\label{csstandardform}
\eeq
All the defining properties of a K\"ahler geometry are automatically
satisfied.

\vspace{0.8cm}
We wish to repeat these steps for the $\s$-model described in terms
of complex linear superfields $\Sigma^a$, $\Sigmab^{\bar{a}}$, 
$a,\bar{a}=1,\dots,n$, satisfying the constraints $\bar{D}^2 \Sigma^a=0$,
$D^2 \Sigmab^{\bar{a}} =0$,
\beq
S=\int d^2x~d^4\theta~K(\Sigma, \Sigmab)
=\frac{1}{4} \int d^2x~d^2\theta_1~d^2\theta_2~K(\Sigma, \Sigmab)
\label{linearaction}
\eeq
Again we expand $\Sigma$ in terms of $N=1$ superfields
\beq
\Sigma(\theta_1,\theta_2)=f(\theta_1) +\frac{1}{2}\theta^\a_2 
\l_\a(\theta_1)+\frac{1}{4}\theta^2_2 G(\theta_1)
\label{Sigmaexpansion}
\eeq
having defined the components as
\beq
f=\Sigma|_{\theta_2=0}\qquad\qquad\l_\a=D_{2\a}\Sigma|_{\theta_2=0}
\qquad\qquad G=-D^2_2\Sigma|_{\theta_2=0}
\label{linearcomponents}
\eeq
(The $N=1$ component expansions are given explicitly in Appendix B).
Now we want to use the constraint that $\Sigma$ satisfies i.e.
 $\Db^2 \Sigma=0$. It can be expressed as follows
\beq
 D^2_2\Sigma=-D^2_1\Sigma+
D^\a_1 D_{2\a} \Sigma
\label{linearconstraint}
\eeq
The situation here is somewhat different as compared to the one in
the chiral example.
In fact, from (\ref{linearcomponents}) and (\ref{linearconstraint}) we see
that in the case of the linear superfield
we can solve only for the highest component in terms of the
lower ones
\beq
G=D^2_1 f - D^\a_1\l_\a
\label {linearsolution}
\eeq
while the $N=1$ superfield $\l_\a$ is not determined. 
Indeed the nonminimal multiplet contains the same physical
degrees of freedom as the minimal one, but it differs from it
in the auxiliary field content. As we will show now the
superfields $\l_\a$ are not dynamical and can be set 
on-shell algebraically.

Using the result in (\ref{linearsolution})
the action (\ref{linearaction}) is
reduced to $N=1$ superspace
\bea
S&=&\frac{1}{8} \int d^2x~d^2\theta_1~\left[  D_1^\a
{\cal{F}}^TMD_{1\a}{\cal{F}}+\L^{\a T}M\L_\a \right. \nonumber\\
&&~~~~~~~~~~\left. -D_1^\a {\cal{F}}^TMP\L_\a
-\L^{\a T}PMD_{1\a}{\cal{F}}\right]
\label{N1linearaction}
\eea
where we have defined
\beq
{\cal{F}}=\left( \matrix{f^a \cr \bar{f}^{\bar{a}}}\right) \qquad \qquad
\L_\a=\left( \matrix{\l_\a^a\cr \bar{\l}_\a^{\bar{a}}}\right)
\eeq
and
\beq
M=\left( \matrix{K_{ab} &K_{a\bar{b}} \cr K_{\bar{a}b}& 
K_{\bar{a}\bar{b}} } \right) \qquad \qquad
P=\left( \matrix{\bf{1} & \bf{0}\cr \bf{0}&-\bf{1}} \right)
\label{defmatrix}
\eeq
We are using the notation
\beq
K_{a\dots \bar{a}\dots}=
\frac{\pa}{\pa f^a}\dots \frac{\pa}{\pa \bar{f}^{\bar{a}}}\dots 
K(f,\bar{f})
\eeq

As anticipated above, 
the superfields $\L_\a$ appear in (\ref{N1linearaction}) 
as auxiliary fields with equations of motion
\beq
\L_\a=M^{-1}PMD_{1\a}{\cal{F}}
\label{linearonshell}
\eeq
Substituting (\ref{linearonshell}) in (\ref{N1linearaction}) we obtain
\beq
S=\frac{1}{8} \int d^2x~d^2\theta_1~\left[ 
D^\a_1 {\cal{F}}^T(M-MPM^{-1}PM)D_{1\a}{\cal{F}} \right]
\label{linearactiononshell}
\eeq
In addition to the manifest invariance under a first
supersymmetry, the action in (\ref{linearactiononshell})
possesses a second one inherited from the $N=2$ reduction 
\beq
\d {\cal{F}}=-\eta^\a M^{-1}PM D_{1\a} {\cal{F}}
\label{N2SSlinear}
\eeq
In general, from dimensional analysis and Lorentz
and parity invariance, a second supersymmetry transformation, 
as in (\ref{complexstructure}), is of the form
\beq
\d {\cal{F}}=i\eta^\a J~D_{1\a}{\cal{F}}
\eeq
where $J$ must be the complex structure of a complex manifold,
as required by the closure of the algebra. 
Thus, from (\ref{N2SSlinear}) we find the expression of the complex structure,
while from the action in (\ref{linearactiononshell}) we determine
the metric
\beq
J=iM^{-1} P M  \qquad \qquad g=M-MPM^{-1} P M 
\label{structuremetric}
\eeq
It is easy to verify that the complex structure $J$ satisfies  
\beq
J^2=-1,\quad\quad J^TgJ=g, 
\quad\quad N_{ij}^{~k} \equiv J_i^{~l}  \pa_{[l} J_{j]}^{~~k}-
 J_j^{~l}  \pa_{[l} J_{i]}^{~~k}=0, \quad\quad \pa_{[k} J_{ij]} =0
\label{kahlerproperties} 
\eeq
where $N_{ij}^{~k}$ is the Nijenhuis tensor.
Thus we conclude that $J$ is the complex structure of a K\"ahler 
manifold described by a nonholomorphic set of coordinates
$(f^a,\bar{f}^{\bar{a}})$. We can go to a canonical basis by
performing the following field redefinition
\beq
{\cal{I}}=\frac{\pa K}{\pa {\cal{F}}} \qquad \qquad 
{\rm{with}} \quad {\cal{I}}\equiv\left( \matrix{\phi^a \cr 
\bar{\phi}^{\bar{a}}} \right) =\left( \matrix{K^a \cr 
K^{\bar{a}}} \right)
\label{fieldredefinition}
\eeq
In terms of the new variables the second supersymmetry
transformation becomes 
\beq
\d {\cal{I}}=M\d {\cal{F}}=M(i\eta^\a J D_{1\a} {\cal{F}})=
i\eta^\a MJM^{-1} D_{1\a}{\cal{I}}
\label{N2SScanonical}
\eeq
so that the new complex structure takes the standard form
\beq
J'\equiv MJM^{-1}=iP
\label{complexcanonical}
\eeq
If we now express the action in  (\ref{linearactiononshell})
using the fields in (\ref{fieldredefinition})
we obtain
\beq
S=\frac{1}{8} \int d^2x~d^2\theta_1~\left[D^\a_1{\cal{I}}^T
(M^{-1}-PM^{-1}P)D_{1\a}{\cal{I}}\right]
\label{N1dualaction}
\eeq
where $M$ has to be thought as a function of $f(\phi,\bar{\phi})$
and $\bar{f}(\phi,\bar{\phi})$ as given by inverting the relation
in (\ref{fieldredefinition}). The matrix $~[M^{-1}-PM^{-1}P]~$
is block off-diagonal so that it consistently represents the 
K\"ahler metric (with only nonvanishing holomorphic-nonholomorphic
components), in complete analogy with the result 
in (\ref{N1chiralaction}) for the chiral formulation. The above
equivalence, however,
has been obtained after the elimination
of the extra auxiliary fields $\l_\a$ via their equations of motion.

We observe that the action (\ref{N1dualaction}) is nothing else than
the $N=1$ formulation we would have obtained
performing a duality transformation on the original action in 
(\ref{linearaction}). Indeed, one can start from the first order action
\beq
S=\int d^2x~d^4\theta\left[ K(\Sigma,\Sigmab)-\Sigma\Phi-
\Sigmab\Phib\right]
\label{firstorderaction}
\eeq
with $\Phi$, $\Phib$ satisfying the chirality constraints
$\Db_\a\Phi=0$, $D_\a \Phib=0$. The $\s$-model in (\ref{linearaction})
is reobtained by functional integration
over $\Phi$, $\Phib$ which imposes the linearity constraints on $\Sigma$,
$\Sigmab$. On the other hand, the equations of  motion for $\Sigma$,
$\Sigmab$ give
\beq
\Phi=\frac{\pa K}{\pa \Sigma}\qquad \qquad \Phib=\frac{\pa K}{\pa \Sigmab}
\eeq
which evaluated at $\theta_2=0$ take the form
\beq
\phi=\frac{\pa K}{\pa f}\qquad \qquad \bar{\phi}=\frac{\pa K}{\pa \bar{f}}
\eeq
i.e. (\ref{fieldredefinition}). In this way one reconstructs a $\s$-model
in terms of chiral superfields with a potential $\tilde{K}$ given 
by the Legendre transform of $K$
\beq
\tilde{K}(\Phi,\Phib)=[K(\Sigma,\Sigmab)-\Sigma\Phi-
\Sigmab\Phib]|_{\Sigma=\Sigma(\Phi,\Phib), \Sigmab=\Sigmab(\Phi,\Phib)}
\label{dualpot}
\eeq
which is just the $N=2$ formulation of the theory in (\ref{N1dualaction}).

The equations of motion (\ref{linearonshell}) for the bosonic components of 
the $N=1$ auxiliary 
superfields are explicitly given in Appendix B.
There one can see that the auxiliary fields $p_{\a\b}$ are expressed in terms
of the space-time derivative of the physical field $B$ and therefore they
acquire a
nontrivial dynamics which modifies the quadratic action
for the field $B$ in a substantial way.      
The same pattern is present in the fermionic 
sector once the equations
(\ref{linearonshell}) are used to eliminate the auxiliary $\bar{\b}_{\a}$
field in terms of derivatives of the physical fermion $\bar{\zeta}_{\a}$.
Thus at the component level, as pointed out in ref. \cite{deogates}, 
the complex linear model seems to differ
from the chiral one. However, since the equations of motion 
for the $\s$-model with linear multiplets are dual to the constraints of 
the chiral one, the two systems must become dual equivalent once the auxiliary
fields of the linear multiplet are set on-shell and correspondingly
the auxiliaries of
the chiral multiplet are eliminated through the constraints.   
Indeed the elimination of the auxiliary fields from 
(\ref{N1linearaction})   
leads to a physical action in $N=1$ superspace (or in components) which 
is the Legendre transform of a chiral action.   

\vspace{0.8cm}

Now we extend our analysis to the general $\s$-model whose potential is
a function of chiral and complex linear superfields \cite{somejim}
\beq
S=\int d^2x~d^4\theta~\O(\Phi^\m,\Phib^{\mb},\Sigma^a,\Sigmab^{\bar{a}})
\label{mixedaction}
\eeq
with $\m,\mb=1,\dots,m$ and $a,\bar{a}=1,\dots,n$. In order to
obtain the $N=1$ reduction we expand the fields as in 
(\ref{Phiexpansion}) and (\ref{Sigmaexpansion}) and
introduce the definitions
\bea
&&H=\left( \matrix{ 0 & \O_{\m \bar{\n}}\cr \O_{\bar{\m} \n} &
0} \right) \qquad \qquad
M=\left( \matrix{\O_{ab} &\O_{a \bar{b}} \cr \O_{\bar{a} b} &
\O_{\bar{a}\bar{b}} }\right) \nonumber\\
&&~~~~~~~~~~~~~\nonumber\\
&& N=\left( \matrix{\O_{a\m} &\O_{a\bar{\m}} \cr \O_{\bar{a} \m} &
\O_{\bar{a}\bar{\m}} }\right) \qquad \qquad
P=\left( \matrix{ \bf{1}& \bf{0}\cr   \bf{0}& -\bf{1}} \right)
\label{mixedmatrix}
\eea
and
\beq
{\cal{I}}=\left(\matrix{\phi^\m\cr \bar{\phi}^{\mb}}\right) \qquad \qquad
{\cal{F}}=\left( \matrix{f^a\cr \bar{f}^{\bar{a}}}\right) \qquad \qquad
\L_\a=\left( \matrix{\l_\a^a\cr \bar{\l}_\a^{\bar{a}}}\right)
\label{mixedvectors}
\eeq
In terms of the above quantities the $N=1$ action takes the form
\bea
&&S=\frac{1}{8}\int d^2x~d^2\theta_1~\left[-2D_1^\a {\cal{I}}^T HD_{1\a}{\cal{I}}
+D_1^\a {\cal{F}}^T MD_{1\a}{\cal{F}} +\L^{\a T}M\L_\a 
-D_1^\a {\cal{F}}^TMP\L_\a \right.\nonumber\\
&&~~~~~~~~~~~\left.  -\L^{\a T}PMD_{1\a}{\cal{F}}~
+\L^{\a T}[N,P]D_{1\a}{\cal{I}}
+D_1^\a {\cal{I}}^T [P,N^T]\L_\a \right]
\label{N1mixedaction}
\eea
Once again the spinor superfields $\L_\a$ are auxiliary and can be
eliminated using the on-shell condition
\beq
\L_\a=M^{-1}PMD_{1\a}{\cal{F}}-M^{-1}[N,P]D_{1\a}{\cal{I}}
\label{mixedonshell}
\eeq
By doing so we find an action in which only the $N=1$
superfields $\phi^\m$,
$\bar{\phi}^{\bar{\m}}$, $f^a$ and $\bar{f}^{\bar{a}}$ appear
and interact in a nontrivial manner
\bea
&&S=\frac{1}{8}\int d^2x~d^2\theta_1~\left[-2D_1^\a {\cal{I}}^T H D_{1\a}{\cal{I}}
+D_1^\a {\cal{F}}^T M D_{1\a}{\cal{F}}  \right.\nonumber\\
&&~~~\left.-(D^\a_1 {\cal{F}}^T M P-D^\a_1 {\cal{I}}^T[P,N^T])
M^{-1}(PMD_{1\a}{\cal{F}}-[N,P]D_{1\a}{\cal{I}})\right]
\label{mixedonshellaction}
\eea
The second supersymmetry transformations are easily computed
\bea
&&\d {\cal{I}}=-\eta^\a PD_{1\a}{\cal{I}} \nonumber\\
&&\d {\cal{F}}=-\eta^\a M^{-1} P M D_{1\a}{\cal{F}}+
\eta^\a M^{-1}[N,P]D_{1\a}{\cal{I}}
\label{mixedSS2}
\eea
They can be rewritten as
\beq
\d {\cal{X}}=i\eta^\a JD_{1\a}{\cal{X}}
\eeq
having defined the vector
\beq
{\cal{X}}=\left(\matrix{{\cal{I}}\cr {\cal{F}}}\right)=
\left(\matrix{\phi^\m\cr
\bar{\phi}^{\bar{\m}}\cr f^a\cr \bar{f}^{\bar{a}}}\right)
\eeq
and the complex structure
\beq
J=i\left(\matrix{P&0\cr ~~~&~~~\cr -M^{-1}[N,P]&M^{-1}PM}\right)
\label{mixedcomplex}
\eeq
It is straightforward to check that $J$ satisfies the conditions
in (\ref{kahlerproperties}) so that the underlying 
manifold is K\"ahler.
In complete analogy with what we had done before, we can go to
a holomorphic basis 
\beq
{\cal{I}}'={\cal{I}}\qquad\qquad {\cal{F}}'=
\left( \matrix{ f^{'a} \cr \bar{f}^{'\bar{a}} }\right )
\label{mixedfieldred}
\eeq
where $f^{'a}\equiv \pa \O/\pa f^a$ and
$\bar{f}^{'\bar{a}}\equiv \pa \O/\pa \bar{f}^{\bar{a}}$.
In the new coordinate system the second supersymmetry transformations
become
\beq
\d {\cal{X}}'=i\eta^\a R J R^{-1} D_{1\a} {\cal{X}}'
\label{SS2mixedcanonical}
\eeq
where
\beq
R=\left(\matrix{1&0\cr N& M}\right)
\eeq
The standard form of the complex structure is then obtained from
(\ref{SS2mixedcanonical})
\beq
J'=R J R^{-1} =i\left(\matrix{P&0\cr 0&P}\right)
\label{mixedcanonical}
\eeq
As a final step one can write the action in terms of the new
fields
\beq
S=\frac{1}{8} \int d^2x~d^2\theta_1~D_1^\a {\cal{X}}^{'T}
{\cal{G}}D_{1\a} {\cal{X}}'
\label{mixedcanonicalaction}
\eeq
and check that the metric ${\cal{G}}$ has only 
barred-unbarred components. 

The above analysis leads to the conclusion that
models described by chiral superfields, complex linear superfields
or both of them, once
reduced to $N=1$ superspace and with
appropriate choices of complex coordinates, i.e. appropriate field
redefinitions, all behave in a very similar manner. In particular
they all share the property of possessing a K\"ahler metric from which
geometrical objects can be easily obtained.
However we wish to emphasize that the reduction procedure for the chiral
case is conceptually different as compared to the complex linear one.
In fact the solution of the chirality constraint allows one to determine
completely the higher $N=1$ components in terms of the lowest one.
On the contrary, for the linear superfield only the highest component 
is given
by the constraint equation, and in order to obtain the 
standard form
for the $N=1$ action one has to use the on-shell conditions.
Thus the relevant question is: are the two formulations, with or 
without auxiliary 
fields equivalent? The results obtained in our previous work \cite{PRVPZ}
seem to confirm the equivalence: there we have  performed a
one-loop calculation using the explicit $N=2$ formalism, and we
have found that the classical duality transformations are
maintained at the quantum level. We study further this issue 
in the next
section where the $\b$-function for the mixed model is computed
at one loop.

\section{The one-loop $\b$-function}

We compute in superspace and
start with the $N=2$ version of the model
\beq
S=\int d^2x~d^4\theta~\O(\Phi^\m,\Phib^{\mb},\Sigma^a,\Sigmab^{\bar{a}})
\label{mixedactionagain}
\eeq
where $\mu, \bar{\mu} = 1, \cdots ,m$ and $a, \bar{a} =1, \cdots ,n$.
In order to perform perturbative calculations it is advantageous
to use the background field method and split the fields 
\beq
\Phi\rightarrow \Phi+\Phi_0\qquad\qquad
\Phib\rightarrow \Phib+\Phib_0\qquad\qquad
\Sigma\rightarrow\Sigma+\Sigma_0\qquad\qquad
\Sigmab\rightarrow\Sigmab+\Sigmab_0
\label{quantumback}
\eeq
The action is then expanded around the background $\Phi_0$,
$\Phib_0$, $\Sigma_0$ and $\Sigmab_0$.
We separate the free kinetic action, which determines the
quantum propagators, from the interactions. We consider
vertices quadratic in the quantum fields since that is all we
need for a one-loop calculation
\bea
&&S=\int d^2x~d^4\theta~\left[ \Phi^\m \Phib^{\mb} \d_{\m\mb}
-\Sigma^a\Sigmab^{\bar{a}} \d_{a\bar{a}}+ (\O_{\m\mb}-\d_{\m\mb})
\Phi^\m\Phib^{\mb} +(\O_{a\bar{a}}+\d_{a\bar{a}}) 
\Sigma^a\Sigmab^{\bar{a}}\right.
\nonumber\\
&&~~~~~~~~~+\frac{1}{2} \O_{\m\n}\Phi^\m\Phi^\n +\frac{1}{2}
\O_{\mb\bar{\n}} \Phib^{\mb} \Phib^{\bar{\n}} +\frac{1}{2}
\O_{ab}\Sigma^a\Sigma^b+\frac{1}{2}\O_{\bar{a}\bar{b}}\Sigmab^{\bar{a}}
\Sigmab^{\bar{b}}\nonumber\\
&&~~~~~~~~~\left.+\O_{a\mb}\Sigma^a \Phib^{\mb}+\O_{\bar{a}\m}\Sigmab^{\bar{a}}
\Phi^\m+\O_{a\m}\Sigma^a \Phi^{\m}+\O_{\bar{a}\mb}\Sigmab^{\bar{a}} 
\Phib^{\mb}+
\dots\right]
\label{quantumaction}
\eea
The quantum fields are explicit while the background is implicit in the
vertices given by derivatives of the potential $\O$. 
Superspace Feynman diagrams and standard D-algebra techniques are the
ingredients for loop calculations. The quantization of the chiral 
superfield is common knowledge, whereas the quantum treatment of
the complex linear superfield has been obtained recently \cite{GVPZ},
\cite{PRVPZ}.
We refer the reader to the relevant references for details. Here we simply
recall that the chiral superspace propagators
are given by
\beq
<\Phib^{\mb}\Phi^\m>=-\frac{\d^{\m\mb}}{\Box}\d^{(4)}(\theta-\theta')
\label{chiralprop}
\eeq
Correspondingly for the complex linear superfield one has an effective
propagator (see \cite{PRVPZ})
\beq
<\Sigmab^{\bar{a}} \Sigma^a>=\d^{a\bar{a}}\left(\frac{D^2\Db^2}{\Box}+
\frac{D_\a\Db^2 D^\a}{\Box}\right) \d^{(4)}(\theta-\theta')
\label{sigmaprop}
\eeq
Additional factors of $\Db^2$ and $D^2$ come from each chiral,
antichiral quantum line respectively, at the vertices. 
We have not mentioned
the infinite tower of ghosts introduced by the Batalin-Vilkovisky
gauge-fixing procedure of the complex linear superfield \cite{GVPZ} 
since, as shown in \cite{PRVPZ}, they essentially
decouple from the external background and do not contribute at one
loop. 

The one-loop $\b$-function is computed evaluating all the
divergent contributions to the effective action: they are given 
by local expressions that by dimensional analysis do not contain
any derivatives. Thus the spinor covariant derivatives always stay
on the quantum lines of the Feynman diagrams and the D-algebra is
straightforward (we make use repeatedly of the identities
listed in Appendix A). Some care is required in collecting
all the
terms with their appropriate combinatoric factors. 
It is convenient to introduce the following notation
\beq
{\cal{W}}\equiv (\O_{\m\mb}-\d_{\m\mb}) \qquad\quad {\cal{V}}\equiv
(\O_{a\bar{a}}+\d_{a\bar{a}}) \qquad\quad {\cal{U}}\equiv\O_{ab}
\qquad\quad{\cal{Z}}\equiv \O_{a\mb}
\label{def}
\eeq
The divergent contributions are then grouped in various sets:
those corresponding to graphs which contain only 
$\Phi {\cal{W}}\Phib$ interactions which give rise to the
one-loop divergence
\beq
\O^{(1)}_1\rightarrow \frac{1}{\e} ~{\rm tr~ ln~}(1+{\cal{W}})
\eeq
We note that, if we were to set to zero the complex linear 
superfields, the above result
would be just the standard one-loop divergence for the $N=2$
chiral $\s$-model with a corresponding metric $\b$-function 
proportional to the Ricci tensor of the K\"ahler manifold
(cfr. eq. (\ref{Ricci})).

Then there are the graphs with $\Sigma{\cal{V}}\bar{\Sigma}$ interactions
which contribute
\beq
\O^{(1)}_2\rightarrow -\frac{1}{\e} ~{\rm tr ~ln}~(1-{\cal{V}})
\label{us1}
\eeq
It is also easy to show that
the sum of diagrams containing any number of $\Sigma{\cal{V}}\bar{\Sigma}$
vertices and an equal number of $\Sigma{\cal{U}}\Sigma$ and
$\bar{\Sigma}\bar{\cal{U}}\bar{\Sigma}$ interactions give rise to a divergent
contribution of the form
\beq
\O^{(1)}_3\rightarrow -\frac{1}{\e} ~{\rm tr~ ln}\left(1-{\cal{U}}
\frac{1}{1-{\cal{V}}}
\bar{{\cal{U}}} \frac{1}{1-{\cal{V}}}\right)
\label{us2}
\eeq
The results in (\ref{us1}) and (\ref{us2}), with chiral superfields 
set equal to zero, have been obtained in ref. \cite{PRVPZ} and lead
to the one-loop $\b$-function for the complex linear $\s$-model.

Finally there are the remaining diagrams which contain both chiral 
and linear quantum lines. In order to account for this type of terms 
let us define an effective chiral propagator with the
${\cal{W}}$-vertices resummed
\beq
<<\Phib \Phi >>=-\frac{1}{\Box}~\frac{1}{1+{\cal{W}}}\d^{(4)}(\theta-\theta')
\equiv \P
\eeq
and an effective linear propagator with the ${\cal{V}}$-vertices and the
${\cal{U}}$- and $\bar{\cal{U}}$-vertices
resummed
\beq
<<\Sigmab \Sigma>>= \left[\frac{D^2\Db^2}{\Box}+
\frac{D_\a\Db^2 D^\a}{\Box} \left( 1-{\cal{U}}\frac{1}{1-{\cal{V}}}
\bar{{\cal{U}}} \frac{1}{1-{\cal{V}}}\right)^{-1}
\right] \frac{1}{1-{\cal{V}}}
\d^{(4)}(\theta-\theta')\equiv \hat{\P}
\eeq
It is easy to show that the last class of diagrams 
can be written in terms of $\P$ and $\hat{\P}$ as
\beq
{\rm tr}\left[{\cal{Z}}\P\bar{{\cal{Z}}}\hat{\P}+\frac{1}{2}
({\cal{Z}}\P\bar{{\cal{Z}}}\hat{\P})^2 +\frac{1}{3}
({\cal{Z}}\P\bar{{\cal{Z}}}\hat{\P})^3+\dots\right]
={\rm tr}\sum_{n=1}^{\infty} \frac{1}{n}({\cal{Z}}\P\bar{{\cal{Z}}}\hat{\P})^n
\eeq
The corresponding divergence is given by
\beq
\O^{(1)}_4\rightarrow \frac{1}{\e} {\rm tr~ ln~}\left(1+{\cal{Z}}\frac{1}
{1+{\cal{W}}}
\bar{{\cal{Z}}} \frac{1}{1-{\cal{V}}}\right)
\eeq
Finally, adding all the various contributions and using (\ref{def}),
we obtain
\beq
\O^{(1)}\rightarrow \frac{1}{\e}\left[ 
{\rm tr~ln~}(\O_{\m\mb}-\O_{\m\bar{a}}\O^{-1}_{\bar{a}a}\O_{a\mb})
-{\rm tr~ln~}(\O_{ab}\O^{-1}_{b\bar{b}}
\O_{\bar{b}\bar{a}}-\O_{a\bar{a}})
\right]
\label{mixeddivergence}
\eeq
Now we show that this result is in perfect agreement with what 
expected from the
duality correspondence between the chiral and the complex linear 
formulations.  Under duality trasformation on the linear superfields
the action in (\ref{mixedactionagain}) is mapped into a
pure chiral $\s$-model
\beq
S_D=\int d^2x~d^4\theta~
\tilde{\O}(\Phi^\m,\Phib^{\mb}, \Psi^a,\bar{\Psi}^{\bar{a}})
\label{chiral2action}
\eeq
where
\beq
\Psi^a=\frac{\pa\O}{\pa\Sigma^a} \qquad \qquad 
\bar{\Psi}^{\bar{a}}=\frac{\pa\O}{\pa\Sigmab^{\bar{a}}}
\label{duality}
\eeq
and $\tilde{\O}$ is the Legendre transform of the $\O$
potential
\beq
\tilde{\O}(\Phi^\m,\Phib^{\mb}, \Psi^a,\bar{\Psi}^{\bar{a}})
=\left[\O(\Phi,\Phib,\Sigma,\Sigmab)-\Psi^a\Sigma^a
-\bar{\Psi}^{\bar{a}}\Sigmab^{\bar{a}}
\right]|_{\Sigma=\Sigma(\Phi,\Phib,\Psi,\bar{\Psi}),
\Sigmab=\Sigmab(\Phi,\Phib,\Psi,\bar{\Psi})}
\label{legendretransform}
\eeq
Our claim is that under duality transformations $\O^{(1)}$
in (\ref{mixeddivergence}) is mapped into the one-loop
divergent contribution that one would obtain from a chiral model
described in terms of coordinates $\Phi^\m, \Psi^a, \Phib^{\mb}, 
\bar{\Psi}^{\bar{a}}$ with $\m,\mb=1,\dots,m$ and $a,\bar{a}=1,\dots,n$
\beq
\O^{(1)}\rightarrow\tilde{\O}^{(1)}=\frac{1}{\e}
{\rm tr~ ln~}\tilde{\O}_{i\bar{j}}
\label{chiraldiv}
\eeq
where now $i,\bar{j}=1,\dots,m+n$.
Let us check this result in the simple case $m=n=1$. (The generalization
is straightforward.) We rewrite (\ref{mixeddivergence}) as
\beq
\O^{(1)}=\frac{1}{\e}{\rm ln~}\frac{\O_{\Phi\Phib}\O_{\Sigma\Sigmab}
-\O_{\Phi\Sigmab}\O_{\Sigma\Phib}}{\O_{\Sigma\Sigma}\O_{\Sigmab\Sigmab}-
\O_{\Sigma\Sigmab}^2}
\label{mixeddiv11}
\eeq
Using the Legendre transform in (\ref{legendretransform}) 
which defines $\tilde{\O}$, we express the derivatives of $\O$
with respect to $\Phi, \Phib, \Sigma, \Sigmab$ in terms of
derivatives of $\tilde{\O}$ with respect to $\Phi, \Phib, 
\Psi, \bar{\Psi}$. With the definition
\beq
\D=\tilde{\O}^2_{\Psi\bar{\Psi}}-\tilde{\O}_{\Psi\Psi}
\tilde{\O}_{\bar{\Psi}\bar{\Psi}}
\label{det}
\eeq
we obtain
\bea
&&\O_{\Sigma\Sigma}=\frac{\tilde{\O}_{\bar{\Psi}\bar{\Psi}}}{\D}
\qquad\qquad \O_{\Sigma\Sigmab}=-\frac{\tilde{\O}_{\Psi\bar{\Psi}}}
{\D} \nonumber\\
&&~~~~~~~~~~~~~~\nonumber\\
&&\O_{\Phi\Sigmab}=\frac{\tilde{\O}_{\Psi\Psi}\tilde{\O}_{\Phi\bar{\Psi}}
-\tilde{\O}_{\Psi\bar{\Psi}}\tilde{\O}_{\Phi\Psi}}{\D}\nonumber\\
&&~~~~~~~~~~~~~\nonumber\\
&&\O_{\Phi\Phib}=\tilde{\O}_{\Phi\Phib}+\frac{1}{\D}
(\tilde{\O}_{\bar{\Psi}\bar{\Psi}}\tilde{\O}_{\Phib\Psi}-
\tilde{\O}_{\Psi\bar{\Psi}}\tilde{\O}_{\Phib\bar{\Psi}})\tilde{\O}_{\Phi\Psi}
\nonumber\\
&&~~~~~~~~
+\frac{1}{\D}(\tilde{\O}_{\Psi\Psi}\tilde{\O}_{\Phib\bar{\Psi}}-
\tilde{\O}_{\Psi\bar{\Psi}}\tilde{\O}_{\Psi\Phib})\tilde{\O}_{\Phi\bar{\Psi}}
\label{inversion}
\eea
Substituting in (\ref{mixeddiv11}) we obtain
\beq
\O^{(1)}=\frac{1}{\e}{\rm ln~}(\tilde{\O}_{\Phi\Phib}\tilde{\O}_{\Psi\bar{\Psi}}
-\tilde{\O}_{\Phi\bar{\Psi}}\tilde{\O}_{\Phib \Psi})
\label{divchiral11}
\eeq
This is indeed the one-loop divergent correction to the effective 
potential of a chiral $\s$-model described by coordinates 
$(\Phi,\Psi,\Phib,\bar{\Psi})$ and K\"ahler potential $\tilde{\O}$
\beq
\tilde{\O}^{(1)}=\frac{1}{\e}~ {\rm tr~ln~}\tilde{\O}_{i\bar{j}}=
\frac{1}{\e}~{\rm ln~ det~}\tilde{\O}_{i\bar{j}}
\eeq
where
\beq
\tilde{\O}_{i\bar{j}}=\left(\matrix{\tilde{\O}_{\Phi\Phib}&
\tilde{\O}_{\Phi\bar{\Psi}}\cr\tilde{\O}_{\Psi\Phib}&
\tilde{\O}_{\Psi\bar{\Psi}}}\right)
\eeq
We note that there is a one-to-one correspondence between
mixed models with vanishing one-loop $\b$-function and chiral 
models with $\b^{(1)}=0$.

In the next section we study the condition of vanishing 
$\b$-function to see whether a theory with $N=4$ supersymmetry
can be constructed.

\section{$N=4$ supersymmetry}

We consider the action in (\ref{mixedactionagain}) in its simplest
version, i.e. $m=n=1$. In this case the target manifold is
$4$-dimensional. With a duality transformation we can map the 
coordinates $(\Phi, \Phib, \Sigma, \Sigmab)$ into a set of chiral 
complex coordinates $(\Phi, \Phib, \Psi, \bar{\Psi})$ and 
correspondingly for the potential
\beq
\O(\Phi,\Phib,\Sigma,\Sigmab) \rightarrow
\tilde{\O}(\Phi,\Phib,\Psi,\bar{\Psi})
\eeq
with $\tilde{\O}$ obtained as Legendre transform of $\O$
as in (\ref{legendretransform}).

It is well known that the chiral model described by the potential
$\tilde{\O}$ has 
$N=4$ extended supersymmetry if the corresponding complex manifold is
hyperka\"hler. In this case, if we express the 
supersymmetry transformations in terms of $N=1$ superfields
\beq
\d \phi^i =i J^{(A)i}_{~~j} \e^{(A)\a} D_\a \phi^j
\qquad\qquad\qquad A=1,2,3
\label{SSN4}
\eeq
the three complex structures must satisfy the standard
conditions (see eq. (\ref{kahlerproperties}))
\beq
J^{(A)2}=-1\qquad\qquad J^{(A)T} gJ^{(A)} = g 
\qquad\qquad N^{(A)k}_{~ij}=0 \qquad \qquad
dJ^{(A)}=0
\label{complexN4}
\eeq
and, in addition, the quaternionic relation
\beq
J^{(A)i}_{~~k}J^{(B)k}_{~~j}=-\d^{AB} \d^{~i}_{j}+
\e^{ABC} J^{(C)i}_{~~j}
\label{quaternionic}
\eeq
We know also that all hyperkahler manifolds are Ricci flat. Thus $N=4$
chiral models have vanishing one-loop $\b$-function, which
from eqs. (\ref{divchiral11}) and (\ref{Ricci}) means
\beq
\tilde{\O}_{\Phi\Phib}\tilde{\O}_{\Psi\bar{\Psi}}
-\tilde{\O}_{\Phi\bar{\Psi}}\tilde{\O}_{\Phib\Psi}
={\rm c-number}
\label{Ricciflat}
\eeq
Since we are considering a 4-dimensional target
manifold, the converse is also true \cite{superspace,hull}, 
i.e. requiring (\ref{Ricciflat})
to be satisfied we have Ricci flatness and consequently the manifold
is hyperkahler and the corresponding chiral $\s$-model is $N=4$
supersymmetric.

Now we want to implement these results directly on the mixed
chiral-linear model. Since duality is maintained at one loop,
a vanishing $\b^{(1)}$-function for the chiral model leads to 
a vanishing $\b^{(1)}$-function for the mixed model. In other 
words the condition in (\ref{Ricciflat}) is dual equivalent to
\beq
\frac{\O_{\Phi\Phib}\O_{\Sigma\Sigmab}
-\O_{\Phi\Sigmab}\O_{\Sigma\Phib}}{\O_{\Sigma\Sigma}\O_{\Sigmab\Sigmab}-
\O_{\Sigma\Sigmab}^2}={\rm c-number}
\label{vanishingbetamixed}
\eeq
which obviously implies $\b^{(1)}_{mixed}=0$, cfr. (\ref{mixeddiv11}).
At this point if we were sure that $N=4$ supersymmetry
is not broken by duality transformations, we could conclude
that a model which satisfies (\ref{vanishingbetamixed})
is $N=4$ supersymmetric. Motivated by this expectation,
we proceed by explicit construction and find a class
of mixed models which exhibit $N=4$ supersymmetry. 

Thus we start with the chiral-linear model in $N=2$ superspace and look 
for two extra supersymmetries, which we demand to mix chiral and linear
multiplets, to be linear in the fields and compatible with
the chiral and linear constraints. These requirements lead to the 
following transformations
\bea
&&\d \Phi=\bar{\e}_\a \Db^\a \Sigma \qquad \qquad \qquad \quad \quad
~\d \Phib=\e_\a D^\a \Sigmab \nonumber\\
&&\d \Sigma=\e_\a D^\a \Phi +\bar{\e}_\a \Db^\a \Sigmab
\qquad \qquad \d \Sigmab=\bar{\e}_\a \Db^\a \Phib +\e_\a D^\a \Sigma
\label{N4transf}
\eea
where $\e_\a=\xi_\a +i \zeta_\a$, $\bar{\e}_\a=\xi_\a-i \zeta_\a$
are the complex parameters. It is easy to show that the 
corresponding algebra closes off-shell
\beq
[\d_{\e^{(1)}},\d_{\e^{(2)}}]=[\d_{\bar{\e}^{(1)}},\d_{\bar{\e}^{(2)}}]=0
\qquad \qquad [\d_\e,\d_{\bar{\e}}]=i\e_\a \bar{\e}_\b \pa^{\a\b}
\label{algebra}
\eeq
Moreover the $\s$-model action is left invariant by the above 
transformations if and only if there exist two functions $G$, 
$\bar{G}$ such that
\bea
&&\O_{\Phib} D^\a \Sigmab+\O_{\Sigma} D^\a \Phi+\O_{\Sigmab} D^\a \Sigma 
=D^\a G \nonumber\\
&&\O_{\Phi}\Db^\a \Sigma+\O_{\Sigmab} \Db^\a \Phib+\O_{\Sigma} 
\Db^\a \Sigmab 
=\Db^\a \bar{G}
\label{N4condition}
\eea
It follows that the second derivatives of the potential must
satisfy
\beq
\O_{\Phi\Phib}=\O_{\Sigma\Sigmab}\qquad\qquad
\O_{\Phi\Sigmab}=\O_{\Sigma\Sigma}\qquad\qquad
\O_{\Sigma\Phib}=\O_{\Sigmab\Sigmab}
\label{condsecondder}
\eeq
The relations in (\ref{condsecondder}) imply (\ref{vanishingbetamixed})
and therefore $\b^{(1)}=0$. However, since the last conditions 
are stronger than (\ref{vanishingbetamixed}) the class of models selected
by (\ref{condsecondder}) does not exhaust the whole class of $N=4$ systems in 
four dimensions. A more general approach would require the definition of
nonlinear supersymmetry transformations, with 
(\ref{vanishingbetamixed}) as integrability  conditions for the invariance
of the action. 

In order to study the geometry underlying the $N=4$ model we
consider its reduction to $N=1$ superspace. We expand the chiral
and the linear superfields into their $N=1$ components as in
(\ref{Phiexpansion}), $(\ref{Sigmaexpansion})$. Moreover
we eliminate the components $F$, $G$, $\psi_\a$  as in
(\ref{chiralsolution}), (\ref{linearsolution}) solving the constraints,
and $\l_\a$ as in (\ref{mixedonshell}) using the on-shell condition.
Finally for the $N=1$ lowest components 
$\chi=(\phi,\bar{\phi}, f,\bar{f})$,  we obtain three extra 
supersymmetries which, as in (\ref{SSN4}), can be expressed 
in the form
\beq
\d \chi= i\eta^\a J^{(1)} D_{1\a} \chi +
i\xi^\a J^{(2)} D_{1\a} \chi +i\zeta^\a J^{(3)} D_{1\a} \chi 
\label{3extraSS}
\eeq
where $J^{(1)}$ is given in (\ref{mixedcomplex}) and
\bea
&&J^{(2)}=\frac{i}{2}\left( \matrix{ PM^{-1}[N,P] &
1-PM^{-1}PM \cr 2+PSM^{-1}[N,P] & S+SPM^{-1}PM} \right)~\nonumber\\
&&~~~~~~~~~~\nonumber\\
&&~~~~~~~~~~\nonumber\\
&&J^{(3)}=\frac{1}{2} \left( \matrix{-M^{-1}[N,P] & -P+M^{-1}PM \cr
2P-SM^{-1}[N,P] & SP+SM^{-1}PM }\right)
\label{2-3complexstructures}
\eea
The matrices $M$, $N$ and $P$ are given in (\ref{mixedmatrix}) and
$S$ is defined as 
\beq
S=\left( \matrix{\bf{0}&\bf{1}\cr \bf{1}&\bf{0}}\right)
\eeq
Through direct computation, one verifies that $J^{(A)}$, 
$A=1,2,3$ satisfy
the conditions in (\ref{complexN4}) and (\ref{quaternionic}), so that 
the complex manifold is hyperka\"hler.

\section{Conclusions}

We have studied two-dimensional supersymmetric $\s$-models
described in terms of {\em both} chiral and complex linear
superfields. Classically the linear superfield is dual
equivalent to the chiral one, therefore on the basis of the 
correspondence $complex ~linear\rightarrow chiral$,
one is naturally lead to borrow the well known results obtained for $N=2$
chiral non linear $\s$-models. The main properties of these 
models are the associated K\"ahler geometry, a one-loop
metric $\b$-function proportional to the Ricci tensor, a
vanishing $\b$-function at two and three loops and a nonvanishing 
correction at four loops \cite{GVZ}. However some caution must be 
used in a straightforward implementation of this program:
one must be aware of the fact that the minimal and the 
nonminimal multiplets do differ in their auxiliary field content.
As we have seen in detail a complete equivalence is obtained only if 
some $N=1$ components of the nonminimal multiplet are set
on-shell. In order to maintain a complete off-shell formulation
of the theory one has to work in $N=2$ superspace. Using this
formalism we have computed the one-loop $\b$-function for the
mixed $\s$-model and shown that it is in perfect agreement with 
the expectations from duality correspondence. 

At this point
it might be interesting to push the calculation at higher 
perturbative orders. In addition since, both chiral and linear 
superfields can interact with supersymmetric
Yang-Mills it would be worth to continue and extend the work started in
ref. \cite{PZ}.

Having at our disposal two $N=2$ multiplets
 that can be consistently coupled and quantized, it is natural
to look for additional supersymmetries. Here we have constructed the
simplest class of models which realizes an $N=4$ invariance with an
underlying hyperka\"hler geometry. This has been achieved starting from the
condition of vanishing $\b$-function and 
considering supersymmetry transformations linear in the fields.
We have restricted our attention to the case of one
minimal scalar coupled to one nonminimal scalar field, but 
the generalization
to the case with $2n+2n$ fields is actually straightforward and can be easily
implemented. It would be interesting to consider the equation $\b^{(1)}=0$
in full generality, allowing for additional supersymmetry transformations
nonlinear in the fields. This would lead to the identification of the most
general class of $N=4$ invariant models.

\medskip
\section*{Acknowledgments.}

\noindent 
This work was
supported by the European Commission TMR program
ERBFMRX-CT96-0045, in which S.P., A.R. and D.Z. are associated 
to the University of Torino and A.S. is associated to K.U. Leuven.
\newpage

\appendix
\sect{Conventions}
In $N=2$ superspace we define the supersymmetry
generators \cite{superspace}
\beq
Q_\a=i(\frac{\pa}{\pa \theta^\a}-\frac{1}{2} \bar{\theta}^\b i\pa_{\a\b})
\qquad \qquad \bar{Q}_\b=
i(\frac{\pa}{\pa\bar{\theta}^\b}-\frac{1}{2} \theta^\a i\pa_{\a\b})
\eeq
which satisfy the algebra
\beq
\{ Q_\a,\bar{Q}_\b\}=i\pa_{\a\b}
\eeq
The spinor covariant derivatives are
\beq
D_\a=\frac{\pa}{\pa \theta^\a}+\frac{1}{2} \bar{\theta}^\b i\pa_{\a\b}
\qquad \qquad \bar{D}_\b=
\frac{\pa}{\pa\bar{\theta}^\b}+\frac{1}{2} \theta^\a i\pa_{\a\b}
\eeq
with
\beq
\{ D_\a,\bar{D}_\b\}=i\pa_{\a\b}
\eeq
We list here some other relations that we have repeatedly used in
the calculation of the $\b$-function:
\beq
[D_\a,\Db^2]=-i\pa_{\a\b}\Db^\b
\eeq

\beq
D^2\Db^2 D^2=\Box D^2
\eeq

\beq
D_\a i\pa^{\a\b} \Db_\b
=-2 D^2\Db^2-D_\a\Db^2 D^\a
\eeq
The $D$-algebra in the loop is completed using the identities
\beq
\d^{(4)}(\theta-\theta') D^2\Db^2\d^{(4)}(\theta-\theta')=
\frac{1}{2}\d^{(4)}(\theta-\theta')D^\a\Db^2 D_\a \d^{(4)}(\theta-\theta')
=\d^{(4)}(\theta-\theta')
\eeq

\vspace{0.8cm}
The reduction to $N=1$ superspace is performed in terms of new
coordinates
\beq
\theta^\a_1=\theta^\a+\bar{\theta}^\a \qquad\qquad
\theta^\a_2=\theta^\a-\bar{\theta}^\a 
\label{newtheta}
\eeq
and corresponding covariant derivatives
\beq
D_{1\a}=D_\a+\Db_\a=2\frac{\pa}{\pa\theta_1^\a}+
\frac{1}{2} \theta_1^\b i\pa_{\a\b}\qquad \qquad
D_{2\a}=D_\a-\Db_\a=2\frac{\pa}{\pa\theta_2^\a}-
\frac{1}{2} \theta_2^\b i\pa_{\a\b}
\eeq
which satisfy $\{D_{1\a}, D_{1\b} \} = 2i \pa_{\a\b}$.
The supersymmetry generators are
\beq
Q_{1\a}=Q_\a+\bar{Q}_\a \qquad \qquad 
Q_{2\a}=Q_\a-\bar{Q}_\a
\eeq

\sect{The nonminimal scalar multiplet in components}
The most general solution to the constraint $\bar{D}^2 \S =0$ has the
following form 
\bea
\S(\theta, \bar{\theta}) &=& 
B + \theta^{\a} \rho_{\a} +\bar{\theta}^{\a} \bar{\zeta}_{\a}
+ \theta^{\a}\bar{\theta}^{\b} (p_{\a \b} -\frac{i}{2} \pa_{\a \b} B)
-\theta^2 H +\theta^2 \bar{\theta}^{\a} \bar{\b}_{\a}  \\ \nonumber
~~&~& +\bar{\theta}^2 \theta^{\a} (-\frac{i}{2} \pa_{\a \b}
\bar{\zeta}^{\b}) - \frac{1}{4} \theta^2 \bar{\theta}^2 (3 \Box B + 2i 
\pa^{\a \b} p_{\a \b})
\label{multiplet}
\ena
where the fields appearing in the expansion are given by 
(we use the conventions of refs. \cite{superspace,deogates})
\bea
&& B = \S| \qquad \qquad \quad \rho_{\a} = D_{\a} \S|  \qquad \quad 
\bar{\zeta}_{\a} = \bar{D}_{\a} \S| \nonumber \\
&& p_{\a \b} = \bar{D}_{\b} D_{\a} \S| \qquad H = D^2 \S| \qquad \quad
\bar{\b}_{\a} = \frac12 D^{\b} \bar{D}_{\a} D_{\b} \S|
\label{components}
\ena
In order to perform the reduction to $N=1$ superspace we rewrite the
previous multiplet as an expansion in $\theta_1$, $\theta_2$ using the
definitions (\ref{newtheta}). In terms of $N=1$ superfields the $\S$ multiplet
is then given by (see eq. (\ref{Sigmaexpansion}))
\beq
\S(\theta_1, \theta_2) = f(\theta_1) + \frac12 \theta_2^{\a} \l_{\a}(\theta_1)
+ \frac14 \theta_2^2 G(\theta_1)
\eeq
where 
\bea
f(\theta_1) &=&  B + \frac12 \theta_1^{\a}(\rho_{\a} + \bar{\zeta}_{\a})
+\frac14 \theta_1^2 (p^{\a}_{~\a} -H)  \nonumber \\
\l_{\a} (\theta_1) &=& (\rho_{\a} - \bar{\zeta}_{\a}) +
\theta_1^{\b} \left( p_{(\a \b)} -\frac{i}{2} \pa_{\a \b} B -
\frac12 C_{\b \a} H \right) -\frac12 \theta_1^2 \left( \bar{\b}_{\a} + 
\frac{i}{2} \pa_{\a \b} \bar{\zeta}^{\b} \right)
\nonumber \\   
G(\theta_1) &=& -(p^{\a}_{~\a}+H) + \theta_1^{\a} \left(  
\bar{\b}_{\a} - \frac{i}{2} \pa_{\a \b} \bar{\zeta}^{\b} \right)
- \frac{1}{4} \theta_1^2 (3 \Box B + 2i \pa^{\a \b} p_{\a \b} ) 
\label{components2}
\ena
Corresponding expansions can be written for the $\bar{\S}$--multiplet 
which satisfies the constraint $D^2 \bar{\S}=0$. Its components are obtained
from the previous ones by simply interchanging barred and unbarred quantities
and $\l_{\a} \rightarrow -\bar{\l}_{\a}$. . 

Using the previous expressions for the $N=1$ superfields we can write 
the equations of motion (\ref{linearonshell}) for the auxiliary fields 
$\L_{\a}$ in components.
For instance, setting the fermions to zero, the equations of motion for 
the auxiliary bosonic fields are
\beq
{\cal P}_{(\a \b)}-\frac12 C_{\b\a} P M^{-1} PM {\cal P}^{\g}_{~\g} = 
\frac{1}{2} (1+PM^{-1}PM) i\pa_{\a\b} {\cal B} + \frac12 C_{\b\a} (1- PM^{-1}PM)
{\cal H}
\label{aux}
\eeq
where we have defined
\beq 
{\cal B} =\left( \matrix{B^a  \cr \bar{B}^{\bar{a}}}
\right) \qquad \qquad 
{\cal P}_{\a\b} =\left( \matrix{p^a_{\a\b}  \cr \bar{p}^{\bar{a}}_{\a\b}}
\right) \qquad \qquad
{\cal H} =\left( \matrix{H^a\cr \bar{H}^{\bar{a}}}\right)
\eeq

\end{document}